# Seeing How Fitting Process Works

*Vera Montalbano and Mauro Sirigu*
*University of Siena, Italy*

*Abstract*
A common problem in teaching Physics in secondary school is due to the gap in terms of difficulty between the physical concepts and the mathematical tools which are necessary to study them quantitatively. Advanced statistical estimators are commonly introduced only a couple of years later than some common physical topics, such as e.g. the electronic circuit analysis. Filling this gap with alternative methods appears to be opportune, in order to let the students reach a full comprehension of the issue they are facing with. In this work we use a smartphone camera and *GeoGebra* to propose a visual method for understanding the physical meaning of a fitting process. The time constant of an RC circuit is estimated by fitting the discharge curve of a capacitor visualized on the screen of an oscilloscope.

## 1     Introduction

In order to engage students in understanding the physical world by constructing and using scientific models to describe, explain, predict, design and control physical phenomena, a central role is played by the fit of data measured in a physics experiment. For developing insight into the structure of scientific knowledge by examining how models fit into theories and for showing how scientific knowledge is validated, students must be engaged in evaluating scientific models through comparison with empirical data (Hestenes, 1997). The discover of a physics law behind the observations needs the definition of a set of parameters, whose number denotes the assumed complexity of the model. In order to investigate and constrain the number and the best estimates of these parameters, a fit of the empirical data is mandatory.

The concept of best fit of a sample of data, how the parameters in the proposed function change to obtain the best fit, the uncertainties related to these parameters are difficult to explain to high school students because of their lacking of advanced mathematical tools. Moreover, many attempts of showing examples can be too abstract or remain obscure because of the use of software which does not show in an explicit way the fitting process. Linear or polynomial regression can be performed, for example, by MS Excel functions or by the Data Tool of the Open Source Physics Project[1]. Anyway, with this kind of approach, the fitting process can be perceived like a kind of obscure black box.

Therefore, it appears to be necessary the development of new teaching techniques which can overcome the gap between usable and required mathematical tools. We propose an example of how this can be easily obtained with a manual parameter optimization by using a Dynamic Geometry Environment (*DGE* in the following) such as *GeoGebra* (Mariotti, 2002).

This effective and active way of teaching how a fitting process works emerged in a course focused on physics lab didactics in pre-service education of physics teachers. The proposal was designed during the data analysis of an experiment on the RC circuit and is therefore intended to be tested with high school students struggling with electronics and electrical engineering.

## 2     The discharge of a capacitor

A typical exercise when dealing with electric circuit analysis is to derive theoretically and measuring experimentally the electric potential discharge curve of a capacitor in an RC circuit. When the circuit is supplied by a square wave generator, in the case that all other electric components are ohmic and for negligible resistance of the cables, the flow of electric charges in the circuit is governed by the first Ohm's Law:

$$-\frac{dQ}{dt}R = \frac{Q}{C} \qquad (1)$$

where $dQ$ is the charge flowing in the circuit within the time $dt$, $Q$ is the charge on each plate of the capacitor of capacity $C$ and $R$ is the resistance of the resistor.

---

[1] http://www.opensourcephysics.org/webdocs/Tools.cfm?t=Datatool

By assuming that at time $t = 0$ the circuit is closed and the capacitor fully charged, the differential equation (1) has solution

$$Q(t) = Q_0 e^{-\frac{t}{\tau}} \qquad (2)$$

where $\tau = RC$ in the time constant of the circuit. By deriving and using again the first Ohm's Law, this equation provides

$$V(t) = V_0 e^{-\frac{t}{\tau}} \qquad (3)$$

which shows the exponential decay of the potential difference between the plates of the capacitor (see Figure 1).

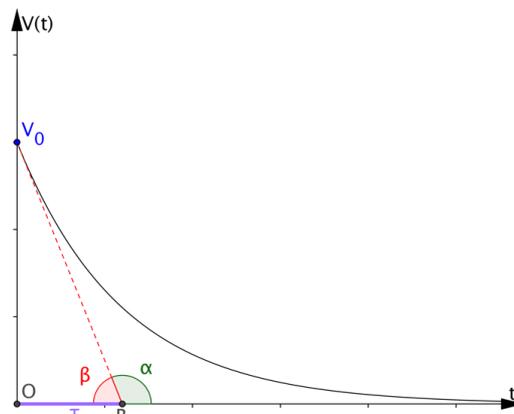

*Fig. 1 - The potential difference between the plate of a capacitor in an RC circuit during a discharge. In purple, the geometrical interpretation of the time constant of the RC circuit.*

By deriving the latter equation and remembering that at time $t$, the potential derivative is graphically represented by the tangent of the angle between the t-axis and the curve $V(t)$ in $t$, we can also show that $\tau$ can be graphically represented as the abscissa of the intersection point between the t-axis and the tangent to the discharge curve.

The time constant is a fundamental parameter in characterizing the circuit response to a signal and can be experimentally measured by sampling the discharge curve of the capacitor.

## 3   Concept and implementation

The use of a DGE for obtaining and processing data is very useful in this context because it allows to achieve multiple goals at the same time. It can: speed up the sampling of the discharge curve; let the students deepen their competences in analytic geometry and show some unexpected applications of it, such as a simple method to understand how fitting process works. Moreover, it can be a captivating way of conducting a physical experiment, since it makes use of multimedia tools, and can also strengthen the students' computer expertness.

### 3.1   GeoGebra

*GeoGebra*[2] is a powerful, easy to learn and open-source software that provides a dynamic geometry environment. This tool is (or should be) extensively used in mathematics teaching (Hall and Chambleeb, 2013; ) to introduce the geometrical construction and mathematical proofs and to provide visual and non-destructive tools for analytic geometry (Hohenwarter and Fuchs, 2005; Baccaglini-Frank and Mariotti, 2010; Leung, Baccaglini-Frank and Mariotti, 2013). It also implements data sheets, advanced statistical analysis tools (Prodromou, 2014) and 3D modeling (Oldknow and Tetlow, 2008). Within the purpose of this work, *GeoGebra* has been used to bypass mathematical difficulties and to obtain a visualisation of the physical meaning of fitting.

---

[2] http://www.geogebra.org

The discharge curve of the capacitor can be easily displayed on the screen of an oscilloscope connected in parallel. With a smartphone it is possible to take a picture of the screen and import it into *GeoGebra*. This procedure allows to sample the curve in a high number of points in very little time: instead of visually taking measures directly on the screen by also taking into account the scaling factor in the time and potential axes, in a few clicks we can shift the picture in *GeoGebra* in order to match the grid of the Cartesian plane and define the units on the *x* and *y-axis*. By simply clicking on the curve, we can create points which will sample it very quickly. We can also define analytically an exponential curve to be superimposed to the picture. The parameters defining this exponential curve can be regulated in order to maximize the match with the experimental one. This procedure defines a manual and visual approach to the fitting process, and also allows an interdisciplinary approach to a physical problem, by implementing analytical geometry tools and also reinforcing computer competences.

A fine tuning of the parameters can also be performed by defining *sliders* as we will better explain in the following.

## 3.2  Materials and methods

As usual in TFA, low budget instruments and electrical components were used in order to reproduce the material commonly available in secondary schools and understand the main experimental difficulties that can be faced during a lab lesson. In particular, the frequency of the square signal from the wave generator and the shape of the exponential discharge curve of the ceramic capacitor can be obtained by sampling the output of an analog oscilloscope (a GWInsteak GOS-622G with a $20 MHz$ band width and a sensitivity of $1 mV/div$ on both channels), added to the circuit as shown in Figure 2. An accurate data reading requires at least a sample of 10 measures of the differential potential as a function of time. Due to the little screen, the spread of the light trace, the scaling factors and the lack of a fine grid, these measures can be in some case a bit difficult.

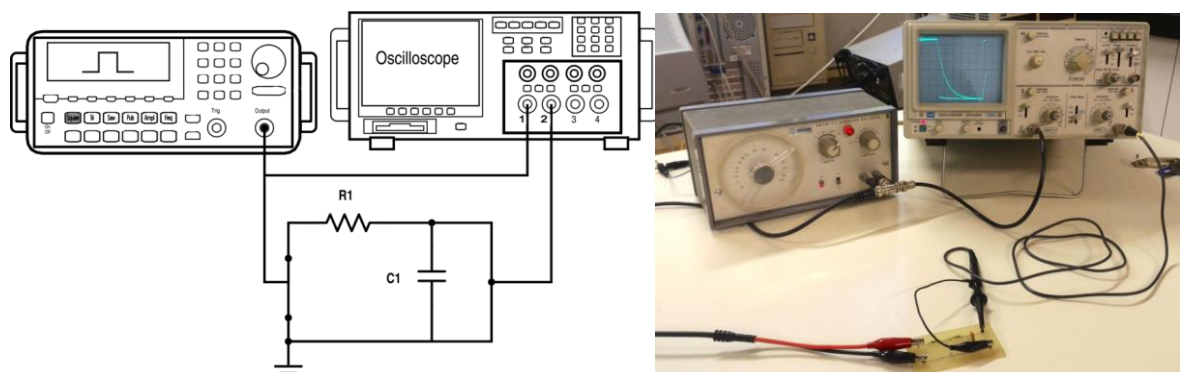

*Fig. 2 - Scheme of the RC circuit with a square wave generator and an oscilloscope in parallel to the capacitor (on the left) and the experimental setup (on the right).*

## 3.3  Procedure

After setting up the circuit as shown in the previous section, we produced a square wave with the generator and visualized it on the screen of the oscilloscope. On the second channel we visualized the discharge curve of the capacitor. We changed the horizontal and vertical shifts and regulated the scaling factors on the *x* and *y*-axes in order to show the discharge curve entirely and as wide as possible, in order to minimize the uncertainty on the sampling.

We took a picture with a smartphone from a distance of about 3 meters, i.e. much greater than the screen size, thus reducing the "fish-eye" effect that could affect the reliability of the measures. The picture was imported into *GeoGebra*, moved and resized to make the oscilloscope grid aligned to the Cartesian axes and the discharge curve begin at t = 0. The sampling of the discharge curve can be done by simply clicking on several points of the curve. The coordinates of the points in the Cartesian plane correspond to the measure of time and potential difference respectively.

Finally, we superimposed to the graph an exponential function defined analytically as in (3) with a free parameter τ. $V_0$ is set to match the starting value of the potential difference just before the start of the discharge phase. The time constant is defined like an Action Object Tool of *GeoGebra*, a Slider tool, which can be changed manually within a certain range and with defined steps. By changing its value, it is possible to visually understand the effects of different time constants on the time response of the circuit.

The τ of the circuit is measured by choosing the value which best matches the experimental curve, as shown in Figure 3. This curve can be compared to the best fit automatically generated by *GeoGebra* with the function *RegExp*, which performs an exponential regression applied to the list of points that sample the discharge curve.

The uncertainty on τ can be evaluated by simply moving the slider and finding the upper and lower approximation to the discharge curve.

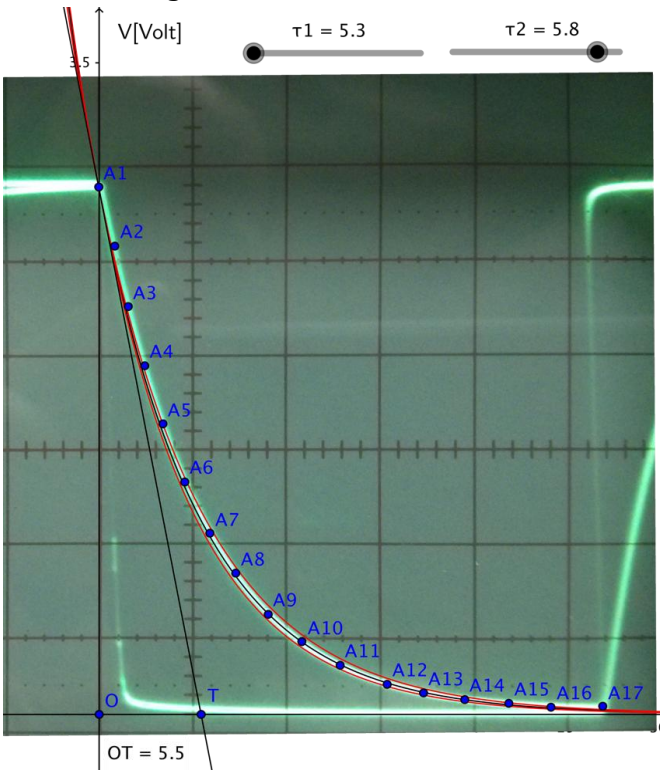

*Fig. 3. Analysis of the discharge curve of a capacitor with GeoGebra. The picture of the discharge curve was imported in GeoGebra and aligned to the grid of the Cartesian plane. The blue points represent the sampling of the curve. The black curve shows the exponential regression of the experimental points, while in red we show the curves corresponding to upper and lower limit to the time constant of the circuit (the respective values are shown by the sliders on top of the image). The black line is the tangent to the best fit curve at $t = 0$ and the length of the segment OT represents the best value for τ in μs.*

Little discrepancies could be noticed between the values of the slider τ and the abscissa of the intersection point between the *x*-axis and the tangent to the discharge curve at $t = 0$. The exploration of the causes of these discrepancies can be an excellent opportunity for a deeper discussion on modelling, fitting and evaluation of uncertainties. Discrepancies could be ascribed to the residual distortion and misalignment of the picture to the grid of the Cartesian plane or to the choice of a model with only one parameter in the fit. In fact, the initial point of the curve was fixed without uncertainties. How this fact can be taken into account in the fitting process and how it changes the final results could be proposed as a further task after the data processing.

## 3.4 Data and Processing

A first measure of the frequency of the square wave can be directly obtained by the generator:
$$\nu_g = (19.00 \pm 0.25) kHz \qquad (4)$$
The semi period of the oscillation is indeed shown also by the screen of the oscilloscope:
$$T/2 = (26.0 \pm 0.5) \mu s \qquad (5)$$
The uncertainty on the frequency is given by
$$\frac{\Delta \nu}{\nu} = \frac{\Delta T}{T} \Rightarrow \Delta \nu = \nu \cdot \frac{\Delta T}{T} = \frac{\Delta T}{T^2} \approx 0.4 kHz \qquad (6)$$
From the semi period we derive $T = (52 \pm 1)\mu s$, which leads to a second measure of the frequency:
$$\nu_o = (19.2 \pm 0.4) kHz \qquad (7)$$
which is consistent with the first one.

The measure of the time constant of the circuit is obtained both manually (by moving the slider to match the experimental discharge curve with the analytic exponential curve) and automatically by the statistical tool of exponential regression provided by *GeoGebra*. The tool returns the best values for the two parameters $A$ and $B$ of a function
$$f(x) = A e^{Bx} \qquad (8)$$
With our sampling points we get $A_* = 2.82V$ and $B_* = -0.19 \mu s^{-1}$, which corresponds to an estimated time constant $\tau = 1/B_* = 5.36 \mu s$.

To obtain an estimate of the uncertainty we define two exponential curves
$$g(x) = A_1 e^{-\frac{x}{\tau_1}} \qquad (9)$$
$$h(x) = A_2 e^{-\frac{x}{\tau_2}} \qquad (10)$$
By moving the sliders $\tau_1$ and $\tau_2$ we find the two curves that envelope the experimental one. We thus get $\tau_1 = 5.1 \mu s$ and t $\tau_2 = 5.6 \mu s$ and define
$$\Delta \tau = \frac{\tau_2 - \tau_1}{2} \qquad (11)$$
We then finally obtain the best estimate for our time constant: $\tau = (5.4 \pm 0.3)\mu s$.

## 4 Remarks and Conclusions

We provided an example of how the use of *GeoGebra* can be of help in sampling and analysing data in a physics experiment. We obtained a measure of the time constant of a RC circuit by importing a picture of the oscilloscope screen in *GeoGebra*, bypassing every difficulty related to the lack of mathematical tools and then comparing the results with the automatic fitting tools included in the software.

The method offers the opportunity to better explain the meaning of mathematical concepts such as the derivatives, the differential equations and the statistical estimators both from a geometric and a physical point of view.

This experiment can also be useful to stimulate curiosity and strategic and critical approach to a physical experiment both reinforcing manual and computer abilities.

The use of *GeoGebra* can show how mathematics and geometry can be used in an unusual way to solve a physical problem.

The example of visualization of the fitting process described in the previous section reported the context in laboratory in which the method emerged and was initially performed. Since the ease and the range of situation in which it can be applied a more effective use of *GeoGebra* in data analysis can be suggested. In the same context, a more sophisticated data fitting could involve the fit of all the parameters of the model.

Moreover, the analysis of electric circuits is not the more effective moment in which introduce the fitting process for at least two reasons. Students are introduced to electricity later in their study in physics, usually after mechanics and thermodynamics. But many experiments in these previous topics can be better understood if the fitting process is already introduced and well-established. Furthermore, the fit of an exponential curve is not the best way for introducing the visualization. Students are more accustomed to manipulating straight lines and interpreting proportional laws. Thus, the next step in progress for implementing the visual fitting is to design a learning path on an issue of mechanics and test it in class.